\documentclass[prl,showkeys,showpacs,groupedaddress,twocolumn]{revtex4-1}
\usepackage{amssymb,comment}
\usepackage{graphicx}
\usepackage{dcolumn}
\usepackage{bm}
\usepackage{amsmath}
\usepackage{subfigure}
\usepackage{float}
\usepackage{color}
\usepackage[colorlinks,linkcolor=blue,anchorcolor=blue,citecolor=blue,bookmarksnumbered]{hyperref}
\newcommand{\PT}{{\cal PT}}
\begin{document}
\title{Experimental Observation of Partial Parity-Time Symmetry and Its Phase Transition with a Laser-Driven Cesium Atomic Gas}
\author{Yongmei Xue$^{1}$}
\thanks{Y. X. and C. H. contributed equally to this work.}
\author{Chao Hang$^{2,3,4}$}
\thanks{Y. X. and C. H. contributed equally to this work.}
\author{Yunhui He$^{1}$}
\author{Zhengyang Bai$^{2}$}
\author{Yuechun Jiao$^{1,4}$}
\author{Guoxiang Huang$^{2,3,4}$}
\thanks{Corresponding author. gxhuang@phy.ecdnu.edu.cn}
\author{Jianming Zhao$^{1,4}$}
\thanks{Corresponding author. zhaojm@sxu.edu.cn}
\author{Suotang Jia$^{1,4}$}
\affiliation{$^{1}$State Key Laboratory of Quantum Optics and Quantum Optics Devices, Institute of Laser Spectroscopy, Shanxi University, Taiyuan 030006, China}
\affiliation{$^{2}$State Key Laboratory of Precision Spectroscopy, East China Normal University, Shanghai 200062, China}
\affiliation{$^{3}$NYU-ECNU Joint Institute of Physics, New York University Shanghai, Shanghai 200062, China}
\affiliation{$^{4}$Collaborative Innovation Center of Extreme Optics, Shanxi University, Taiyuan, Shanxi 030006, China}

\date{\today}

\begin{abstract}
Realization and manipulation of parity-time ($\PT$) symmetry in multidimensional systems are highly desirable for exploring nontrivial physics and uncovering exotic phenomena in non-Hermitian systems. Here, we report the first experimental observation of partial $\PT$ (p$\PT$) symmetry in a cesium atomic gas coupled with laser fields, where a two-dimensional p$\PT$-symmetric optical potential for probe laser beam is created.
A transition of the p$\PT$ symmetry from an unbroken phase to a broken one is observed through changing the beam-waist ratio of the control and probe laser beams, and the domains of unbroken, broken, and non-p$\PT$ phases are also discriminated unambiguously. Moreover, we develop a technique to precisely determine the location of the exceptional point of the p$\PT$ symmetry breaking by measuring the asymmetry degree of the probe-beam intensity distribution.
The findings reported here pave the way for controlling multidimensional laser beams in non-Hermitian systems via laser-induced atomic coherence, and have potential applications for designing new types of light amplifiers and attenuators.
\end{abstract}

\maketitle

{\it Introduction.--} Although non-Hermitian Hamiltonians generally have complex eigenvalues, a wide class of non-Hermitian Hamiltonians with parity-time ($\PT$) symmetry was found to exhibit all-real spectra~\cite{Bender1998,Bender2005,Bender2007,Konotop,Ganainy}. The exploration of $\PT$-symmetry has provided an excellent platform for uncovering the exotic behaviors in systems with open environments and spawned intriguing prospects of non-Hermitian physics~\cite{Ashida2020}.
The study of $\PT$-symmetric Hamiltonians was soon introduced into optics~\cite{Ganainy2007,Musslimani2008,Makris2008,Feng2017} based on a close analogy between the Schr\"{o}dinger equation in quantum mechanics and the Maxwell equation under paraxial approximation in electrodynamics.
An optical $\PT$ symmetry can be created when the potential $V(\bf r)$ controlling light propagation is made to be invariant under complex conjugation and simultaneous reflection in all spatial directions, i.e. $V^*({\bf r})=V(-{\bf r})$. In recent years, there have been intensive studies on the realization of optical $\PT$ symmetry in various physical settings, including waveguide and fiber arrays~\cite{Guo2009,Ruter2010,Szameit}, photonic circuits and lattices~\cite{Feng2011,Regensburger},
microtoroid resonators~\cite{Peng2014,Chang2014,Wen2018}, trapped ions~\cite{Ding2021}, etc. Compared to other materials, atomic media are desirable for realizing non-Hermitian Hamiltonians by synthesizing desired refractive index profiles due to their nice coherence property and the superiority for available active manipulations on light absorption, gain, dispersion, and nonlinearity, and so on~\cite{Hang2013,Sheng2013,Peng2016,Zhang2016,Wen2019,Jiang2019,Hang2017,
ZhangXiao2018}.

Note that to have an all-real spectrum for a non-Hermitian Hamiltonian, the condition of $\PT$ symmetry is neither sufficient nor necessary. When non-Hermiticity increases, the spectrum of a $\PT$-symmetric Hamiltonian will become complex, indicating that a phase transition occurs from unbroken $\PT$ phase to broken $\PT$ phase. The transition point between the unbroken and broken phases is called exceptional point (EP), at which both eigenstates and eigenenergies coalesce~\cite{Heiss2012,Miri2019}. So far, many interesting properties and promising applications have been found for $\PT$-symmetric systems, such as Bloch oscillations~\cite{Longhi2009,Wimmer2005}, nonreciprocal and unidirectional invisible light propagations~\cite{Ramezani2010,Feng2011,Lin2011}, coherent perfect absorbers~\cite{Longhi2010,Chong2011,Sun2014,Hang2016}, giant light amplification~\cite{Konotop2012}, single-mode lasers~\cite{Feng2014,Hodaei2014}, phonon lasers~\cite{Jing2014,Zhang2018}, topological energy transfer~\cite{Xu2016}, mode switching~\cite{Doppler2016}, enhanced sensing~\cite{Hodaei2014,Chen2017}, asymmetric diffraction~\cite{Liu2017,Shui2018}, quantum information~\cite{Xiao2017,Xiao2020}, and quantum state tomography~\cite{Naghiloo}, etc.

Recently, a class of multidimensional potentials invariant under complex conjugation and reflection in only one direction [e.g. $V^\ast(x,y)=V(-x,y)$ or
$V^\ast(x,y)=V(x,-y)$ in two dimensions (2D)], called as {\it partial} $\PT$ (p$\PT$) symmetric potentials, have been found theoretically to support all-real spectra. Interestingly, systems with such potentials display also phase transitions from the unbroken phase to the broken one~\cite{Yang2014,Katarshov2015}. The study of the p$\PT$ symmetry can provide a way for realizing multidimensional potentials with all-real spectra without requiring strict $\PT$ symmetry in high dimensions.

In this Letter, we report the first experimental observation of p$\PT$ symmetry and its phase transition. We consider a cesium atomic vapor driven by a probe and a control laser beams, both are half overlapped by a repumping laser beam, resulting in gain and loss for the probe beam and hence the production of a 2D p$\PT$-symmetric potential for the propagation of the probe beam.
Furthermore, a transition of the p$\PT$ symmetry from an unbroken phase to a broken one is observed through adjusting the beam-waist ratio of the control and probe beams; the domains of unbroken, broken, and non-p$\PT$ phases are also discriminated clearly.
In addition, a technique for precisely determining the location of the EP of the p$\PT$ symmetry breaking is developed through the measurement of the asymmetry degree of the probe intensity distribution.
The theoretical analysis and numerical simulation based on Maxwell-Bloch equations are also carried out, which reproduce the all experimental observations well.
The results on the p$\PT$ symmetry and its phase transition reported here open a route for actively manipulating multidimensional laser beams in non-Hermitian systems, and have potential applications for designing new light amplifiers and attenuators.

{\it Experimental setup.--} The schematic of the experimental setup and the related atomic excitation scheme is shown in Fig.~1.
The experiment is performed with a room temperature cesium vapor cell of length 2~cm and diameter 2.5~cm. A weak probe and a strong control beams  (with half Rabi frequencies $\Omega_p$ and $\Omega_c$~\cite{note1}, respectively) are overlapped and co-propagate through the cell; see Fig.~1(b). They drive the atomic transitions $|1\rangle\rightarrow|3\rangle$
and $|2\rangle\rightarrow|3\rangle$, respectively, with detunings given by
$\Delta_3=\omega_p-(E_3-E_1)/\hbar$ and $\Delta_2=\omega_p+\omega_c-(E_2-E_1)/\hbar$, where $E_{\alpha}$ is the eigenenergy of the atomic level $|\alpha\rangle$ ($\alpha=1,\,2,\,3$) [see Fig.~1(a)].

Both the probe and the control beams are of Gaussian profiles and their $1/e^{2}$ waists are $w_p$ and $w_c$, respectively. The beam-waist ratio of the control and probe beams is $\sigma \equiv w_c/w_p$, which is varied between 1.5 and 5 by tuning the waist of the control beam while keeping $w_p = 110~\mu$m. The repumping laser (half Rabi frequency $\Omega_{r}$) has an elliptical Gaussian profile with $1/e^{2}$ waist $w_{r1} = 200~\mu$m ($w_{r2} = 700~\mu$m) for short (long) axis. It counter-propagates through the cell and covers the half of probe and control beams [see the inset of Fig.~1(b)]. In this way, a 2D optical potential with the p$\PT$ symmetry [i.e. $V^*(x,y)=V(-x,y)$] for the probe propagation can form.
After passing through the vapor cell, the probe beam is detected by a photodiode (PD) for monitoring its absorption and detected by a charge coupled device (CCD) for attaining its intensity distribution. The atomic density ${\cal N}_a$ can be changed by varying the temperature of the cell placed in a thermal chamber (for more details, see the supplemental material~\cite{SM}).
\begin{figure}[htbp]
\centering
\includegraphics[width=1\columnwidth]{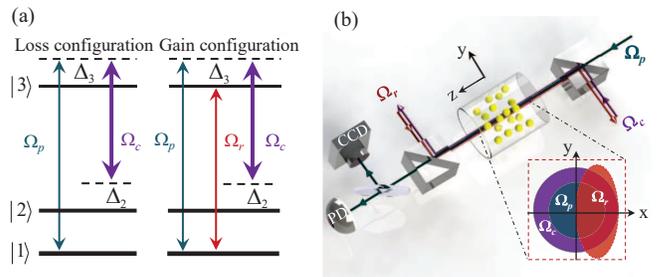}
\caption{
Experimental design for observing the p$\PT$ symmetry and its phase transition.
(a)~Three-level excitation scheme of cesium atoms.
A weak probe laser ($\Omega_{p}$) couples the transition $|1\rangle\rightarrow|3\rangle$ and a strong control laser ($\Omega_{c}$) couples $|2\rangle\rightarrow|3\rangle$, forming the loss configuration in the left panel, with $\Delta_j$ ($j=2$, 3) frequency detunings. A repumping laser ($\Omega_{r}$) resonantly drives the transition $|1\rangle\rightarrow|3\rangle$, forming the gain configuration in the right panel.
(b)~Sketch of the experimental setup. The control beam is fully overlapped with the probe beam, both of them are Gaussian and co-propagate through the cell. The repumping beam (with an elliptical Gaussian profile) counter-propagates with the probe and control beams and covers the right-half region of the probe beam (see  the inset), which creates a  2D p$\PT$-symmetric potential for the probe beam. The output probe beam is detected with PD and CCD camera.
}
\end{figure}

{\it Observation of the p$\PT$ symmetry and its phase transition.-- }
In the experiment, we lock the frequencies of the three laser beams such that
$\Delta_2=0$ and $\Delta_3=2\pi\times 607$\,MHz, and the frequency of the repumping laser is resonant with
$|1\rangle\rightarrow|3\rangle$. In the absence of the repumping beam, the probe beam experiences a loss (i.e. an absorption), which gives a PD signal $S_L$; in the presence of the repumping beam it experiences respectively a gain and a loss in the right- and left-half regions,
which gives a PD signal $S_{GL}$. When the gain and loss are exactly balanced, the PD signal $S_{GL}$ would be zero.

\begin{figure*}[htbp]
\centering
\includegraphics[width=1.9\columnwidth]{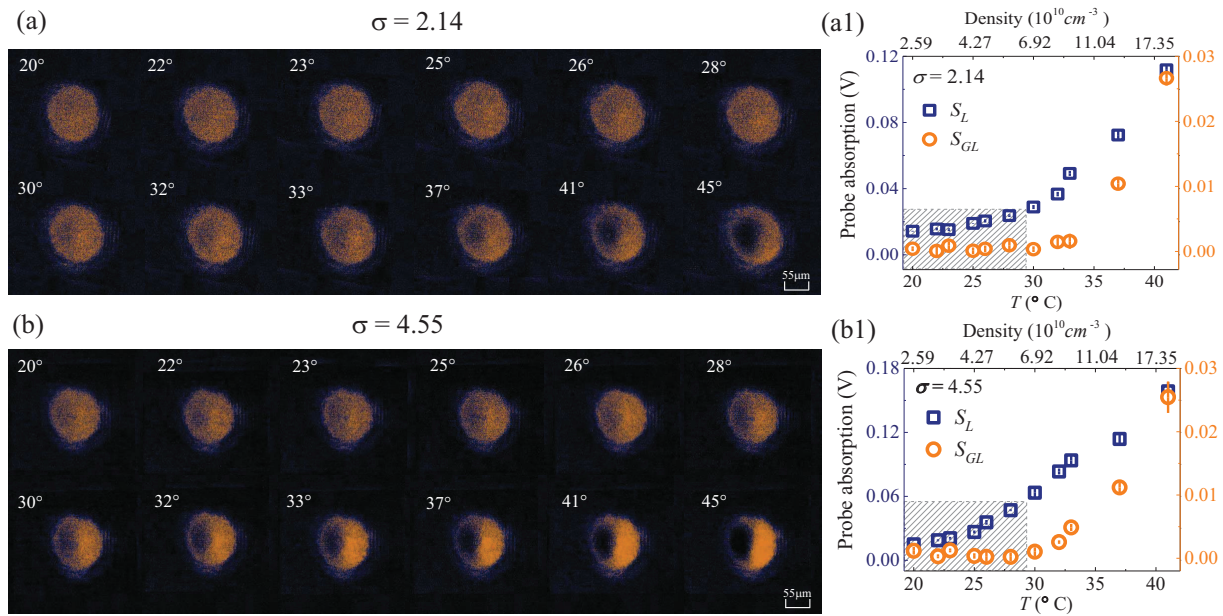}
\caption{Measurements of the p$\PT$ symmetry and its phase transition obtained by changing the atomic-cell temperature and the beam-waist ratio $\sigma \equiv w_c/w_p$ between the control and probe beams.
(a)~The measured result of the probe intensity distribution for $\sigma = 2.14$. The uniform (un-uniform) distribution observed for $T < 29^{\circ}$C  ($\gtrsim 29^{\circ}$C) indicates that the system works in a regime of the p$\PT$ symmetry (non-p$\PT$ symmetry). No p$\PT$  phase transition occurs in this case.
(a1)~Measured result of the probe absorption as a function of temperature $T$ (atomic density ${\cal N}_a$) with (orange circles) and without (blue squares) the repumping beam.
$S_{GL}$ ($S_{L}$): the PD signal in the presence (absence) of the repumping beam. Shadow square: the domain where the system works with the p$\PT$ symmetry.
(b) and (b1) display similar measurements to (a) and (a1) but for $\sigma= 4.55$; in this situation, the probe beam displays an un-uniform intensity distribution for all temperatures due to the p$\PT$ symmetry and its phase transition into a broken p$\PT$ symmetry (see the text for details).
}
\end{figure*}

Shown in the upper part of Fig.~2(a) is the measured result of the probe intensity distribution from the CCD in the presence of the repumping beam for $\sigma = 2.14$.
The intensity distribution is uniform (un-uniform) when $T<$ 29$^\circ$C ($T\gtrsim 29^{\circ}$C). The degree of asymmetry of the distribution for $T\gtrsim 29^{\circ}$C is increased as $T$ increases, which becomes the most evident at $T$ = 45$^\circ$C, i.e. the intensity displays clearly half-dark (left) and half-bright (right) distribution. The orange circles shown in Fig.~2(a1) are results of the probe absorption (i.e. signal $S_{GL}$) measured from the PD as a function of $T$ (${\cal N}_a$). We see that $S_{GL}$ keeps nearly zero for $T< 29^{\circ}$C (or ${\cal N}_a< 6.29\times10^{10}$~cm$^{-3}$), indicating that the gain and loss are balanced and hence the system works in a regime of p$\PT$ symmetry (marked by the shadow square domain in the figure); $S_{GL}$ begins to increase for $T\gtrsim 29^{\circ}$C, indicating that the gain-loss balance is lost and thus un-uniform intensity distribution occurs [corresponding to the lower row of Fig.~2(a)]. Note that this is not due to the breaking of the p$\PT$ symmetry (i.e. not phase transition), but due to the non-p$\PT$ symmetry resulted from the growth of the spontaneous emission of $|3\rangle$ as temperature increases. For comparison, the probe absorption in the absence of the repumping beam (i.e. signal $S_{L}$ from PD) is also given by the blue squares in Fig.~2(a1), which increases with $T$ (${\cal N}_a$), indicating that the probe beam always suffers a significant loss and hence no p$\PT$ symmetry occurs.

In order to observe not only the p$\PT$ symmetry but also its phase transition in the system, we take $\sigma$ as a tunable parameter and make new measurements. Plotted in Fig.~2(b) and (b1) are measured results similar to Fig.~2(a) and (a1) but for $\sigma$ = 4.55. We see that in this situation the probe beam displays asymmetric intensity distributions for all temperatures [Fig.~2(b)]. Meanwhile, $S_{\rm GL} \approx 0$ for $T<$ 29$^\circ$C, meaning that the gain-loss balance is kept and the system works in a p$\PT$-symmetric phase [the first row in Fig.~2(b) and the shadow domain in Fig.~2(b1)]. Consequently, the non-uniformity of the probe intensity distribution for $T<$ 29$^\circ$C [the first row of Fig.~2(b)] {\it must} be the outcome by p$\PT$-symmetry breaking, i.e. the system has entered into a broken p$\PT$-symmetric phase from the unbroken p$\PT$-symmetric phase.

The above experimental findings can be analyzed quantitatively by defining the asymmetry degree of the probe intensity distribution
\begin{align}
\label{eq_D}
D_{\rm asym}=(I_{p,\,{\rm right}}-I_{p,\,{\rm left}})/(I_{p,\,{\rm right}}+I_{p,\,{\rm left}}),
\end{align}
where $I_{p,\,{\rm left}}$ ($I_{p,\,{\rm right}}$) is the average of the probe intensity in the left-half (right-half) part of the distribution, with $D_{\rm asym}=0$ and $D_{\rm asym}\in(0,1]$ characterizing
uniform and un-uniform intensity distributions, respectively.

Measured (samples) and fitted (lines) results of $D_{\rm asym}$ as a function of $T$  for $\sigma$ = 1.70, 2.14, 3.63, 3.95 and 4.55 are shown in Fig.~3(a), respectively.
\begin{figure*}[htbp]
\centering
\includegraphics[width=1.95\columnwidth]{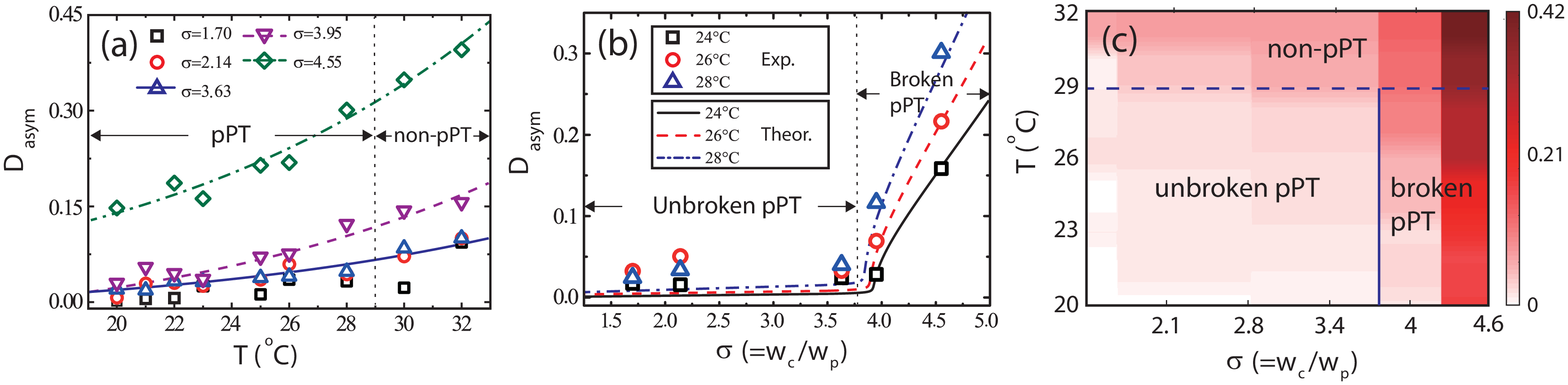}
\caption{Discrimination of the unbroken, broken, and non-p$\PT$ phases.
(a)~Measurements (samples) and fittings (lines) of the asymmetry degree $D_{\rm asym}$ as a function of temperature $T$ for beam-waist ratio $\sigma$ = 1.70, 2.14, 3.63, 3.95 and 4.55, respectively.
(b)~Measurements (samples) and calculations (lines) of $D_{\rm asym}$ as a function of $\sigma$, respectively for T = $24^{\circ}$C, $26^{\circ}$C, and $28^{\circ}$C, with the EP locating at $\sigma=\sigma_{cr}\simeq3.8$.
(c)~Phase diagram by taking $D_{\rm asym}$ as a function of $T$ and $\sigma$, where domains of the unbroken, broken, and non-p$\PT$ phases are shown. The solid (dashed) line indicates the boundary between domains of the unbroken p$\PT$-symmetric phase and the broken one.
}
\end{figure*}
We see that $D_{\rm asym}$ increases slowly with $T$; however, it increases abruptly from 0 when $\sigma$ exceeds a critical value $\sigma_{cr}$. Illustrated in Fig.~3(b) are measured (samples) and calculated (lines) results of $D_{\rm asym}$ as a function of $\sigma$, respectively for $T = 24^{\circ}$C, $26^{\circ}$C, and $28^{\circ}$C. It reveals clearly that a phase transition of p$\PT$ symmetry indeed occurs, with the EP locating at $\sigma=\sigma_{cr}\simeq3.8$. Based on Fig.~3(a) and (b), a phase diagram is obtained by taking $D_{\rm asym}$ as a function of $T$ and $\sigma$ in Fig.~3(c), where domains of the unbroken, broken, and non-p$\PT$ phases are displayed unambiguously. The solid (dashed) line indicates the boundary between domains of the unbroken p$\PT$-symmetric phase and the broken one.

{\it Theoretical analysis and numerical simulation.--}
The Maxwell-Bloch equations describing the probe propagation
are solved by using a perturbation method for $\Omega_p\ll\Omega_c,\,\Omega_r$. Gain-loss property of the probe beam can be obtained from the linear dispersion relation of the system (see \cite{SM} for more details).

According to the experiment, the control-beam intensity has the form $I_c(\xi,\eta)$ $\approx$ $I_{c0}[1-(\xi^2+\eta^2)/\sigma^2]$, with  $(\xi,\eta)$ = $(x,y)/w_p$ and $I_{c0}$ the maximum intensity~\cite{note1}. The propagation equation of the probe beam takes the form  $i\partial_\zeta \Omega_p=-d (\partial_{\xi\xi}+\partial_{\eta\eta}) \Omega_p -V(\xi,\eta) \Omega_p$, with $\zeta$ = $z/L$; $L$ is the cell length, $d= L/L_{\rm diff}$ with $L_{\rm diff}$ = $2k_pw_p^2$  the characteristic diffraction length. The potential in the equation, $V(\xi,\eta)$, can be written in the form  $V(\xi,\eta)=V_1(\xi)+V_2(\eta)$. Here $V_1(\xi)=L(\beta_{G} \xi^2 + i\gamma_{G})$ and $V_2(\eta)=L\beta_{G} \eta^2$ [$V_1(\xi)=L(\beta_{L} \xi^2 + i\gamma_{L})$ and $V_2(\eta)=L\beta_{L} \eta^2$] for $\xi>0$ ($\xi<0$), where $\beta_{G,L}$ = $\sigma^{2}I_{c0}$\,Re$(\partial k_{G,L}/\partial I_c)|_{I_c=I_{c0}}$ and $\gamma_{G,L}$ = Im$(k_{G,L}-\omega/c)|_{I_c=I_{c0}}$, with $k_{G}$ ($k_{L}$) the linear dispersion relation with (without) the repumping laser. Thus, once the condition
\begin{align}
\beta_{G}=\beta_{L}=\beta,\quad \gamma_{G}=-\gamma_{L}=\gamma,
\end{align}
is fulfilled, one has $V(\xi,\eta)=V^\ast(-\xi,\eta)$, i.e. $V(\xi,\eta)$ is p$\PT$-symmetric potential.

From the above analysis we have the following conclusions (which are in agreement with the experimental results given above):
(i)~The location of EP of the p$\PT$ symmetry is determined by the ratio between the loss and gain, i.e. $\gamma/\beta$, which is {\rm not} dependent on the atomic density ${\cal N}_a$. Thus, it is not available to observe the breaking of the p$\PT$ symmetry through increasing the atomic density (the cell temperature). (ii)~The ratio $\gamma/\beta$ is proportional to $\sigma^2$. Therefore, one can observe the breaking of the p$\PT$ symmetry by increasing $\sigma$ in the system.

\begin{figure}[htbp]
\centering
\includegraphics[width=1\columnwidth]{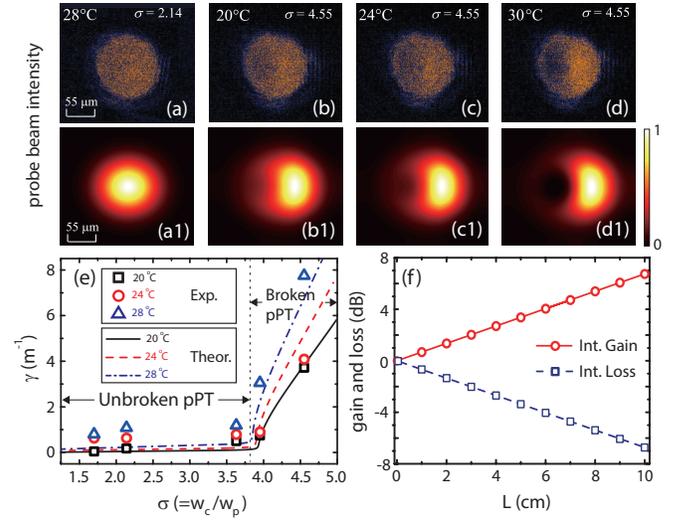}
\caption{The control of the p$\PT$ symmetry.
(a)-(d)~Measured result [for different $(\sigma,T)$] of the probe intensity distribution, which is uniform in (a) due to the unbroken p$\PT$ symmetry, un-uniform in (b) and (c) due to the broken p$\PT$ symmetry, and un-uniform in (d) due to the non-p$\PT$ symmetry.
(a1)-(d1)~Numerical results corresponding to (a)-(d).
(e)~Measured (samples) and calculated (lines) results of the gain and loss coefficient $\gamma$ as a function of $\sigma$ for different $T$.
(f)~Red circles (blue squares): the gain and loss of output probe intensity observed in the left (right) part of the probe beam as a function of the cell length $L$ for $(\sigma,T)$=$(4.55, 28^{\circ}$C). The gain (loss) of the probe beam in the right (left) part can arrive 7 dB (-7 dB) at $L=10$ cm. }
\end{figure}

For a further comparison between theory and experiment, the upper part of Fig.~4 shows
the probe intensity distribution for $(\sigma,T)=(2.14,28^{\circ}$C), $(4.55,20^{\circ}$C), $(4.55,24^{\circ}$C), and $(4.55,30^{\circ}$C), respectively. The first (second) row is the result given by experiment (theory).
The distribution is uniform in (a) and (a1) due to the perfect p${\PT}$ symmetry, un-uniform in (b), (b1) and (c), (c1) due to the breaking of the p$\PT$ symmetry, and un-uniform in (d) and (d1) due to the non-p$\PT$ symmetry. We see that the theory agrees with the experiment well.

{\it Applications for light amplifier and attenuator.--}
The relation between the gain-loss coefficient $\gamma$ and the asymmetry degree of the probe intensity $D_{\rm asym}$ is given by
\begin{equation}
\gamma=\ln [(1+D_{\rm asym})/(1-D_{\rm asym})]/(4L).
\end{equation}
Since the measurement of $D_{\rm asym}$ can reach a high precision (the relative standard deviation $\lesssim$5\%), we can determine the location of EP rather precisely. Fig.~4(e) shows a measurement (samples) and a simulation (lines) on $\gamma$ as a function of $\sigma$ for $T = 20^{\circ}$C, $24^{\circ}$C, and $28^{\circ}$C, respectively. Similar to Fig.~3(b), the mutation of $\gamma$ clearly reveals the breaking of p$\PT$ symmetry with the location of EP ($\sigma_{cr}\simeq3.8$). Fig.~4(f) shows the output probe intensities respectively in the right (gain) and left (loss) parts as functions of $L$; one sees that for a 10-cm-long cell with $(\sigma,T)$ = $(4.55,28^{\circ}C)$, the increase (decrease) of the probe intensity in the right (left) part can arrive 7 dB (-7 dB). Therefore, the present system is promising for designing new types of optical devices that can realize a light amplifier and attenuator in different parts of a single laser beam.

{\it Conclusion.--} We have carried out, for the first time, the experimental observation on p$\PT$ symmetry by using a laser-driven cesium atomic gas; the transition of the p$\PT$ symmetry from an unbroken phase to a broken one has been measured; the unbroken, broken, and non-p$\PT$ phases are discriminated clearly. We have also developed a technique to precisely determine the location of the EP of the p$\PT$ symmetry breaking. The experimental results have been verified well by theoretical calculations. Our work paves the way for controlling multidimensional laser beams in non-Hermitian optical systems, and have potential applications for designing new types of light amplifiers and attenuators.

J. Z. is supported by the National Key R\&D Program of China (Grant No. 2017YFA0304203), the National Natural Science Foundation of China (Grant Nos. 61835007, 11434007, 61775124, and 11804202), Changjiang Scholars and Innovative Research Team University of Ministry of Education of China (Grant No.~IRT\_17R70) and 1331KSC. G. H., C. H., and Z. B. are supported by the National Natural Science Foundation of China (Grant Nos.~11975098, 11974117, and 11904104). C. H. is also supported by the National Key R\&D Program of China (Grant Nos.~2016YFA0302103 and 2017YFA0304201), and Shanghai Municipal Science and Technology Major Project (Grant No.~2019SHZDZX01).


\begin{thebibliography}{99}
\bibitem{note1}To realize the optical potential with  p$\PT$ symmetry, $\Omega_c$ is taken to $(x,y)$-dependent (see \cite{SM}).
\end{thebibliography}
\end{document}